\documentclass[%
 aps,
showtitles,
 amsmath,amssymb,
preprint,pre
]{revtex4-1}

\usepackage{xurl}
\usepackage{graphicx}% Include figure files
\usepackage{dcolumn}% Align table columns on decimal point
\usepackage{bm}% bold math
\usepackage[mathlines]{lineno}% Enable numbering of text and display math
\usepackage{siunitx}
\usepackage{amsmath}
\usepackage[utf8]{inputenc}
\usepackage[T1]{fontenc}
\usepackage{mathptmx}
\usepackage{etoolbox}
\usepackage[table]{xcolor}

\usepackage{float}
\usepackage{graphicx}
\usepackage{dcolumn}
\usepackage{bm}
\usepackage{color}
\usepackage{amsmath}
\usepackage{amsthm}
\usepackage{amssymb}
\usepackage[cal=cm,scr=euler]{mathalpha}

\makeatletter
\def\@email#1#2{%
 \endgroup
 \patchcmd{\titleblock@produce}
  {\frontmatter@RRAPformat}
  {\frontmatter@RRAPformat{\produce@RRAP{*#1\href{mailto:#2}{#2}}}\frontmatter@RRAPformat}
  {}{}
}%
\makeatother

\begin{document} 
\title{Thermodynamics of coupled electromagnetic and mass fields}
\author{Fei Wang}

\email{fei.wang@kit.edu; ORCID: 0000-0003-3318-3264}
\affiliation{Institute of Nanotechnology (INT),
Karlsruhe Institute of Technology (KIT),\\
Hermann-von-Helmholtz-Platz 1, 76344 Eggenstein-Leopoldshafen, Germany}

 \date{\today}

\begin{abstract}
We formulate a unified thermodynamic framework for chemical, mechanical, and electromagnetic transport by extending the Gibbs--Duhem relation to include kinetic free-energy contributions associated with material and field motion. The resulting kinetic Gibbs--Duhem relation establishes a thermodynamic connection between free-energy variations, generalized chemical potentials, pressure, and transport momentum, while preserving the invariance of the system free energy under changes of reference frame. Within a Lagrangian formulation of free-energy evolution, the relation provides a common structure for diffusion, momentum transport, and electromagnetic dynamics. For nonuniform chemical systems, the framework recovers the coupling between Cahn--Hilliard dynamics and the Korteweg stress arising from composition gradients. Its extension to electromagnetic systems incorporates electric, magnetic, charge, and mechanical contributions into the same thermodynamic formulation, yielding a unified description of electromagnetic and momentum evolution. The framework further separates dissipative from nondissipative transport and establishes a common thermodynamic form for mechanical and electromagnetic propagation. In the isentropic limit, the characteristic propagation velocity is determined by the derivative of the thermodynamic pressure with respect to density, yielding the speed of sound for mechanical systems and the speed of light for electromagnetic systems. These results show that kinetic contributions can be incorporated consistently into the Gibbs--Duhem framework, providing a common thermodynamic basis for coupled chemical, mechanical, and electromagnetic transport without introducing separate thermodynamic principles for each mechanism.

\end{abstract}

\maketitle

\section{Introduction}

Thermodynamics provides a fundamental framework for describing macroscopic systems through relations among free energy, chemical potentials, pressure, entropy, and other state variables. Among its central results, the Gibbs--Duhem relation constrains the thermodynamic variables of an equilibrium system and establishes a direct connection between free-energy descriptions and continuum thermodynamics~\cite{gibbs1928collected,sprik2025thermodynamics}. The interpretation of entropy, which connects macroscopic thermodynamic behavior with microscopic degrees of freedom, has likewise been considered from both thermodynamic and statistical perspectives, including its relation to molecular configurations and symmetry~\cite{popp2025flame,bormashenko2020entropy}. These connections motivate energetic and variational formulations in which macroscopic transport can be related systematically to an underlying thermodynamic structure. Such formulations have been extended to nonequilibrium transport~\cite{deGroot1984NonEquilibrium,huang1987statistical,maugin1999thermodynamics}, phase transitions, and complex fluids, where free-energy dissipation provides a systematic basis for deriving evolution equations~\cite{wang2024wetting}. For nonuniform systems, generalized Gibbs--Duhem relations have also been developed within phase-field formulations, establishing connections between chemical potentials and mechanical quantities such as osmotic stress~\cite{alhasan2022gibbs}. These developments motivate a broader question: can the Gibbs--Duhem relation be extended to incorporate the kinetic contributions associated with transport, and can the resulting thermodynamic structure provide a common description of material and electromagnetic transport?

For material systems, free-energy formulations based on dissipation naturally lead to evolution equations for composition and momentum. In a Lagrangian description, the total derivative follows the motion of a material element and therefore contains both local and convective contributions. This distinction becomes important when kinetic energy is incorporated into the thermodynamic description. As demonstrated in the first series of this work~\cite{wang2026pof}, the differential of the kinetic free-energy contribution can be decomposed into a dissipative term associated with the evolution of the transport velocity and a nondissipative term that constrains the corresponding momentum and pressure. This observation suggests that transport momentum can be incorporated into the Gibbs--Duhem framework on the same thermodynamic footing as conventional state variables.

The purpose of the second series of this work is to develop this construction systematically. We formulate free-energy evolution within a Lagrangian framework in which the differential of the free energy is separated into dissipative contributions, which determine the evolution equations, and nondissipative contributions, which satisfy a Gibbs--Duhem-type constraint. The central result is the kinetic Gibbs--Duhem relation, which extends the conventional Gibbs--Duhem relation by incorporating the kinetic contribution associated with material transport~\cite{wang2026pof},
\begin{equation}
\tau\mathrm d\mu_\tau+c\mathrm d\mu+\mathrm dp
+\mathbf u\cdot\mathrm d(\rho\mathbf u)=0.
\label{eq:11-00}
\end{equation}
Here, $\tau$ is the temperature, $c$ is the composition, $p$ is the pressure, $\rho$ is the material mass density, $\mathbf u$ is the velocity field, and $\mu_\tau$ and $\mu$ are the thermal and chemical potentials, respectively. The final term represents the kinetic contribution and provides the corresponding momentum constraint. Thus, within the Lagrangian formulation, diffusion and momentum transport emerge from complementary contributions to the same free-energy differential.

We first apply this framework to nonuniform chemical systems. In classical van der Waals and Cahn--Hilliard descriptions~\cite{rowlinson1979translation,cahn1958free,cai2024chemo}, the free energy contains a composition-gradient contribution that accounts for the energetic cost of spatial inhomogeneity. The thermodynamic interpretation of this gradient contribution, particularly its connection to entropy and molecular-scale motion, is not explicit in the conventional formulation. Within the present framework, a composition gradient enters directly into the kinetic Gibbs--Duhem relation. The resulting momentum balance establishes a connection between the Lagrangian thermodynamic formulation, Cahn--Hilliard dynamics, and the Korteweg stress associated with diffuse interfaces. Gradient-induced mechanical forces can therefore be incorporated into the same thermodynamic structure that governs chemical diffusion and momentum transport.

We next extend the formulation to electromagnetic transport. The electromagnetic free energy is expressed in terms of electric, magnetic, charge, and material-transport contributions, leading to corresponding Gibbs--Duhem relations for electromagnetic-field evolution and electromagnetic--mechanical coupling. These relations provide a common thermodynamic structure for the associated field and momentum equations and establish a direct connection between electromagnetic transport and the kinetic Gibbs--Duhem framework. In appropriate limits, the formulation recovers the conventional Maxwell equations while retaining the Lagrangian, or convective, contributions associated with material transport. The same structure also provides a thermodynamic description of nondissipative electromagnetic transport, including the Meissner state in type-I superconductors and the absence of a transverse force on vortices in type-II superconductors. In addition, the Lagrangian formulation retains coupling terms associated with the curl components of the electromagnetic fields, yielding corresponding contributions to the momentum balance.

Finally, the common structure of the momentum balances for electromagnetic and material fields provides a thermodynamic connection between electromagnetic and mechanical propagation. Under isentropic conditions, the pressure response determines the corresponding characteristic propagation velocity. Acoustic and electromagnetic propagation can therefore be described within a common thermodynamic structure, with their characteristic velocities determined by the same pressure--density relation rather than by separate thermodynamic principles.

The remainder of this paper is organized as follows. Section II introduces the physical setup, notation, and Lagrangian description used throughout the work. Section III develops the kinetic Gibbs--Duhem relation for a single transported particle in both material and electromagnetic systems and establishes its connection with momentum balance, Galilean invariance, and Lorentz invariance. Section IV applies the formulation to uniform and nonuniform chemical systems and connects the resulting momentum balance with Cahn--Hilliard dynamics and the Korteweg stress. Section V extends the free-energy formulation to electromagnetic fields and derives the corresponding Gibbs--Duhem relations and momentum balance. Section VI summarizes the main results and discusses possible extensions of the framework.

\section{Setup and Notations}

\subsection{Physical setup}

We consider a binary system consisting of a charged particle and a surrounding medium within a closed domain $V$ bounded by $\Gamma$. The local state of the particle--medium system is characterized by the mass density $\rho$, volume composition field $c$, and charge density $\rho_e$. An electromagnetic field is present throughout the domain, with electric and magnetic field intensities denoted by $\mathbf E$ and $\mathbf H$, respectively. This continuum description, in which material and field variables are treated as coupled field quantities, follows the general framework of continuum field theories~\cite{eringen1999microcontinuum}. The velocity field associated with mass and charge transport is denoted by $\mathbf u$, while the propagation velocity of the electromagnetic field is denoted by $\mathbf c$. The electric current density is defined as
\begin{equation}
\mathbf J=\rho_e\mathbf u.
\end{equation}

\subsection{Lagrangian and Eulerian frameworks}

We adopt a Lagrangian description for the scalar field $c(\mathbf r(t),t)$ and the vector field $\mathbf u(\mathbf r(t),t)$, where the position $\mathbf r(t)$ follows the motion of a material element. The transport velocity is therefore defined by
\begin{equation}
\mathbf u=\frac{\mathrm d\mathbf r}{\mathrm dt}.
\label{eq:10-9}
\end{equation}
The position vector can be expressed as
\begin{equation}
\mathbf r(t)=\sum_{i=1}^{n}x_i(t)\mathbf e_i,
\label{eq:17-9999}
\end{equation}
where $\mathbf e_i$ is the unit vector in the $i$th coordinate direction and $x_i$ is the corresponding coordinate. Substitution into Eq.~\eqref{eq:10-9} gives
\begin{equation}
\mathbf u=\sum_i\dot{x}_i\mathbf e_i
=\sum_i u_i\mathbf e_i,
\label{eq:10-90}
\end{equation}
where $u_i$ is the velocity component in the $i$th direction.

The total time derivatives of the scalar and vector fields along the material trajectory are
\begin{align}
\frac{\mathrm dc}{\mathrm dt}
&=\frac{\partial c}{\partial t}
+\mathbf u\cdot\nabla c,\\
\frac{\mathrm d\mathbf u}{\mathrm dt}
&=\frac{\partial\mathbf u}{\partial t}
+\mathbf u\cdot\nabla\mathbf u.
\end{align}
Thus, the Lagrangian derivative contains both the local temporal variation and the convective contribution associated with material transport.

In contrast, the Eulerian description represents the fields as functions of fixed spatial position and time, $c(\mathbf r,t)$ and $\mathbf u(\mathbf r,t)$. In this description, $\mathbf r$ and $t$ are independent variables, and the velocity field is specified at each spatial location and time rather than obtained directly from the trajectory of a material element. The distinction between Lagrangian and Eulerian descriptions is fundamental in
continuum mechanics, with the former following material elements and the latter
describing fields at fixed spatial locations~\cite{hutter2004fluid}.

\subsection{Incompressibility condition}

We next clarify the definition of incompressibility in the Lagrangian framework. In the Lagrangian description, the trajectory of a material volume element $\delta V$ is followed. An incompressible material element is characterized by the invariance of its volume,
\begin{equation}
\frac{\mathrm d(\delta V)}{\mathrm dt}=0.
\label{eq:6-88}
\end{equation}
If the mass of each material element is conserved, this condition is equivalent to
\begin{equation}
\frac{\mathrm d\rho}{\mathrm dt}=0.
\end{equation}
When mass exchange between neighboring material elements is allowed, however, volume conservation and density conservation are no longer equivalent. A material element may retain its volume while its density changes through mass transfer, for example, by diffusion. Thus, in systems with mass diffusion, incompressibility is more appropriately defined by the invariance of the material volume rather than by the constancy of the local density.

For each volume element, the volume fraction of the $i$-th chemical component is defined as
\begin{equation}
c_i=\frac{\delta V_i}{\delta V}.
\end{equation}

\section{A single particle}

\subsection{Mass field and Galilei invariance}

We first consider a single massive particle, such as an electron, moving with velocity $\mathbf{u}(\mathbf{r}(t),t)$ and mass $m(\mathbf{r}(t),t)$. The kinetic free energy is written as
\begin{equation}
E_k=m\mathbf{u}^2.
\end{equation}
Because the velocity depends on the reference frame, this kinetic contribution alone is not invariant under a Galilean transformation. To incorporate the constraint associated with the kinetic contribution and retain a nonzero mass or volume associated with the particle, we introduce a Lagrange multiplier $P$ into the free energy~\cite{blokhuisYoungsLawGravity1995a,gibbs1928collected,johnsonConflictsGibbsianThermodynamics1959},
\begin{equation}
E=m\mathbf{u}^2-P.
\end{equation}
For $m=0$, the corresponding Lagrange multiplier also vanishes. Rewriting the kinetic contribution as $m\mathbf{u}\cdot\mathbf{u}$, its total differential can be expressed as
\begin{equation}
\mathrm dE
=m\mathbf{u}\cdot\mathrm d\mathbf{u}
+\mathbf{u}\cdot\mathrm d(m\mathbf{u})
-\mathrm dP.
\label{eq:10-9989}
\end{equation}
The first term represents the dissipative contribution associated with the evolution of the transport velocity and the corresponding entropy production,
\begin{equation}
m\mathbf{u}\cdot\frac{\mathrm d\mathbf{u}}{\mathrm dt}
=-\frac{\mathrm ds_u}{\mathrm dt}.
\end{equation}

The standard Gibbs--Duhem relation can be obtained by separating the dissipative contribution from the remaining part of the free-energy differential. Following the same construction, the remaining terms in Eq.~\eqref{eq:10-9989} give a Gibbs--Duhem-type relation that constrains the mechanical contribution,
\begin{equation}
\boxed{
\mathbf{u}\cdot\mathrm d(m\mathbf{u})-\mathrm dP=0.
}
\label{eq:10-5}
\end{equation}
Integrating this relation gives
\begin{equation}
P=m\mathbf{u}^2.
\end{equation}

Dividing Eq.~\eqref{eq:10-5} by $\mathrm dx_i$ gives
\begin{equation}
\frac{\mathrm d\mathbf r}{\mathrm dx_i}\cdot
\frac{\mathrm d(m\mathbf u)}{\mathrm dt}
=-\frac{\mathrm d\mathcal P}{\mathrm dx_i},
\qquad
\mathcal P=-P.
\label{eq:12-7}
\end{equation}
Since $\mathrm d\mathbf r/\mathrm dx_i=\mathbf e_i$ and
$\mathbf e_i\cdot\mathrm d(m\mathbf u)=\mathrm d(mu_i)$, Eq.~\eqref{eq:12-7} can be written in vector form as
\begin{equation}
\frac{\mathrm d(m\mathbf u)}{\mathrm dt}
=-\nabla\mathcal P,
\end{equation}
which is the momentum balance equation.

The resulting modified free energy satisfies
\begin{equation}
E=m\mathbf u^2-P=0,
\end{equation}
as a consequence of the kinetic Gibbs--Duhem relation. Within the present formulation, the Lagrange multiplier therefore removes the frame-dependent contribution of the kinetic energy and provides a thermodynamic constraint associated with Galilean invariance. The momentum balance can consequently be interpreted as the mechanical expression of this thermodynamic constraint. This perspective differs from formulations based directly on conservation laws, such as those underlying Noether's theorem and the Boltzmann transport equation~\cite{huang1987statistical}.

\subsection{Electromagnetic field and Lorentz invariance}

The above derivation can be extended naturally to the transport of electromagnetic fields, represented here by propagation with a constant velocity $\mathbf c$. For electromagnetic-field transport, the entropy production vanishes, giving
\begin{equation}
m\mathbf c\cdot\frac{\mathrm d\mathbf c}{\mathrm dt}
=-\frac{\mathrm ds_c}{\mathrm dt}=0
\quad\Rightarrow\quad
\frac{\mathrm d\mathbf c}{\mathrm dt}=\mathbf 0.
\end{equation}
Although the propagation velocity $\mathbf c$ is constant, the kinetic contribution $E_k=m\mathbf c^2$ depends on the observer through the Lorentz transformation because the effective transport mass $m(\mathbf r(t),t)$ associated with the electromagnetic field changes between reference frames. In particular, the observed frequency $\omega$ of electromagnetic radiation depends on the observer, and the corresponding photon energy $\hbar\omega$ changes accordingly ($\hbar$: reduced Planck constant). Since $\hbar\omega=m\mathbf c^2$ within the present formulation, the effective mass $m$ changes between reference frames, consistent with the Doppler shift of electromagnetic radiation~\cite{roi2024doppler}, including the red- and blue-shift phenomena.

To account for this frame dependence, we introduce a Lagrange multiplier in analogy with the material-particle formulation. The modified energy is then constrained by the Gibbs--Duhem relation according to
\begin{equation}
m\mathbf c^2-P=0,
\end{equation}
which is invariant under the corresponding Lorentz transformation within the present formulation. The resulting relation,
\begin{equation}
P=m\mathbf c^2,
\end{equation}
has the same form as the familiar mass--energy relation and provides the thermodynamic basis for extending the Gibbs--Duhem framework from material transport to electromagnetic-field transport.

In the following sections, we extend the Gibbs--Duhem formulation to chemical systems, electromagnetic fields, and coupled chemical--electromagnetic systems.

\section{Chemical system}

\subsection{Uniform chemical system}

The free energy functional of a uniform chemical system is written as 
\begin{equation}
 \mathcal E=\int_V [f(\tau,c) +  \rho \mathbf{u}^2+ p] \delta V=\int_V (e+ p) \delta V,
 \label{eq:1-00}
\end{equation}
where $f(\tau, c)$ is a Helmholtz-type free energy due to thermal motion of molecules/atoms, which can be expressed as 
\begin{equation}
 f(\tau, c)=\rho_0 \mathbf{v}^2,
\end{equation}
where $\rho_0$ is the mass density of the molecules/atoms
and  $\mathbf{v}$ denotes the microscopic mean velocity.
In the absence of microscopic motion, corresponding to the zero-temperature limit, both the entropy and its associated free-energy contribution vanish.
$\rho \mathbf{u}^2$ depicts the kinetic free energy of macroscopic motion,
and $p$ is a Lagrange multiplier for  the constant volume element $\delta V$.
The free energy density $e$ is defined as $e=f+ \rho \mathbf{u}^2$.

Applying the standard thermodynamics based on Euler's homogeneous function theorem to the chemical free energy,
the total derivative of the free energy functional reads~\cite{deGroot1984NonEquilibrium}
\begin{equation}
 \mathrm d  \mathcal E=\int_V[\mu_\tau \mathrm d\tau + \tau\mathrm d\mu_\tau+   \mu \mathrm d c+ c\mathrm d \mu+ \rho \mathbf{u}\cdot \mathrm d \mathbf{u}+ \mathbf{u}\cdot \mathrm d (\rho \mathbf{u})+\mathrm d p]\delta V,
 \label{eq:18-88}
\end{equation}
where $\mu_\tau=\partial e/\partial \tau$
and $\mu=\partial e/\partial c$ are respectively the thermal and chemical potentials.
The first, third, and fifth terms depict the diffusion equations
for the temperature $\tau$,  the composition $c$, and the velocity $\mathbf{u}$, respectively.
\begin{align}
\label{eq:20-1}
 \frac{\mathrm d \tau}{\mathrm d  t}&=\nabla \cdot D_\tau\nabla \mu_\tau,\\
 \label{eq:20-2}
 \frac{\mathrm d c}{\mathrm d  t}&=\nabla \cdot D_c\nabla \mu,\\
  \label{eq:20-3}
  \frac{\mathrm d \mathbf{u}}{\mathrm d  t}&=\nabla \cdot D_u\nabla \boldsymbol \mu_u.
\end{align}
The velocity potential $ \boldsymbol \mu_u$ is defined as $\boldsymbol \mu_u=\frac{1}{2}\partial e/\partial \mathbf{u}$.
The mobilities,
$D_\tau$, $D_c$, and $ D_u$
are the coefficients determining the entropy production rate.
The entropy production follows
\begin{align}
 \mu_\tau\frac{\mathrm d \tau}{\mathrm d  t}&=-\frac{\mathrm d  s_\tau}{\mathrm d t},\\
 \mu\frac{\mathrm d c}{\mathrm d  t}&=-\frac{\mathrm d  s_c}{\mathrm d t},\\
 \boldsymbol  \mu_u\cdot \frac{\mathrm d \mathbf{u}}{\mathrm d  t}&=-\frac{\mathrm d  s_u}{\mathrm d t}.
\end{align}

The rest terms in Eq.~\eqref{eq:18-88}  are the kinetic Gibbs-Duhem relation with the chemical and thermal effects:
\begin{equation}
\boxed{
 \tau\mathrm d\mu_\tau+c\mathrm d \mu+ \mathrm d p+\mathbf{u}\cdot \mathrm d (\rho \mathbf{u})=0.
 }
\end{equation}
The first three terms are the standard Gibbs-Duhem relation
and the last term is the new contribution by considering the dissipation of the kinetic free energy.
The kinetic Gibbs-Duhem relation can be readily transformed into the momentum balance equation
with chemical and thermal effects:
\begin{equation}
 \frac{\mathrm d (\rho \mathbf{u})}{\mathrm d t} =-\nabla p-c\nabla \mu-\tau\nabla  \mu_\tau.
\end{equation}

\subsection{Nonuniform chemical system}

The limitations of local thermodynamics in describing capillary phenomena have been discussed extensively by Rowlinson and Widom, motivating the introduction of gradient-dependent free-energy functionals for inhomogeneous fluids~\cite{rowlinson1982molecular}. For a binary, nonuniform system, we write the free-energy functional as
\begin{equation}
\mathcal E=\int_V
\left[
\kappa(\nabla c)^2+\rho\mathbf u^2+p
\right]\delta V,
\label{eq:1-0}
\end{equation}
where $\kappa$ is the gradient-energy coefficient associated with the van der Waals square-gradient formulation~\cite{rowlinson1979translation,cahn1958free}.

We interpret the gradient contribution in Eq.~\eqref{eq:1-0} as an effective microscopic kinetic-energy density associated with molecular motion induced by spatial variations in composition. Specifically, we write
\begin{equation}
f(c,\nabla c)
=\rho_0\mathbf v_{\mathrm{mix}}^2
=\rho_0(\mathbf v_1c_1+\mathbf v_2c_2)^2
=\rho_0c_1^2\mathbf v_1^2
+\rho_0c_2^2\mathbf v_2^2
+\rho_0c_1c_2\mathbf v_1\cdot\mathbf v_2,
\end{equation}
where the first two terms together correspond to $f(\tau,c)$ and describe the mechanical mixture associated with a homogeneous system. The cross term represents the deviation of the free energy from this mechanical-mixture state and is identified here with the gradient contribution to the free energy. 
In general, $\rho_0$ may also depend on the composition; allowing for such a dependence does not alter the present interpretation of the gradient energy.

Within this interpretation, the nonlocal gradient energy originates from the coexistence of two molecular species with distinct microscopic velocities. These microscopic velocities should be distinguished from the macroscopic transport velocity $\mathbf u$ and the corresponding macroscopic mass density $\rho$. By contrast, the term $\rho\mathbf u^2$ represents the macroscopic kinetic-energy contribution associated with continuum transport. The local energy density can therefore be written schematically as
\begin{equation}
e=f+\kappa(\nabla c)^2
+\rho\mathbf u^2+p.
\end{equation}
The Lagrange multiplier $p$ enforces the corresponding thermodynamic constraint on the combined microscopic and macroscopic contributions, since neither \(\mathbf{u}\) nor \(\mathbf{v}\) is Galilean invariant.

This interpretation provides a possible microscopic basis for the gradient-energy term and suggests that molecular-scale motion, its spatial variation, and macroscopic transport can be incorporated within a unified Lagrangian thermodynamic framework.

The thermodynamic contribution of entropy to the gradient energy is not explicitly identified in the conventional van der Waals--Cahn formulation. To examine this contribution, we write the gradient energy as $\kappa\nabla c\cdot\nabla c$. The total differential of the free-energy functional can then be expressed as
\begin{equation}
\mathrm d\mathcal E
=\int_V
\left[
\kappa\nabla c\cdot\mathrm d\nabla c
+\rho\mathbf u\cdot\mathrm d\mathbf u
+\nabla c\cdot\mathrm d(\kappa\nabla c)
+\mathbf u\cdot\mathrm d(\rho\mathbf u)
+\mathrm dp
\right]\delta V.
\label{eq:15-50}
\end{equation}
The first term represents the dissipative contribution associated with the evolution of the composition gradient and the corresponding entropy production. We describe this evolution by
\begin{equation}
\frac{\mathrm d\nabla c}{\mathrm dt}
=\nabla\cdot D_c\nabla(\nabla c).
\end{equation}
The second term gives the velocity-diffusion contribution discussed above. The remaining terms therefore constitute the Gibbs--Duhem relation including the gradient-energy contribution,
\begin{equation}
\boxed{
\nabla c\cdot\mathrm d(\kappa\nabla c)
+\mathbf u\cdot\mathrm d(\rho\mathbf u)
+\mathrm dp=0.
}
\label{eq:23-50}
\end{equation}
Integration of this relation gives
\begin{equation}
-p=\rho\mathbf u^2+\kappa(\nabla c)^2.
\label{eq:37-8}
\end{equation}

Dividing Eq.~\eqref{eq:23-50} by $\mathrm dx_i$ and collecting the spatial components yields the momentum balance
\begin{equation}
\frac{\mathrm d(\rho\mathbf u)}{\mathrm dt}
=-\nabla p-\nabla c\cdot\nabla(\kappa\nabla c).
\end{equation}
Thus, the composition gradient contributes directly to the momentum balance through the same kinetic Gibbs--Duhem structure that governs the uniform system.

To connect this Lagrangian formulation with the conventional Eulerian description, we expand the total derivative of the composition gradient. The contribution of the gradient energy to the free-energy evolution is
\begin{align} 
\frac{\mathrm d\mathcal E}{\mathrm dt} =\int_V \kappa\nabla c\cdot \frac{\mathrm d\nabla c}{\mathrm dt} \delta V =\int_V \kappa\nabla c\cdot \left( \frac{\partial\nabla c}{\partial t} +\mathbf u\cdot\nabla\nabla c \right)\delta V. 
\end{align}
Applying integration by parts to the first term gives
\begin{align} \frac{\mathrm d \mathcal E}{\mathrm d t }=\int_V \bigg\{-\nabla \cdot (\kappa \nabla c)\frac{\mathrm d c}{\mathrm d t } + [\nabla\cdot (\kappa\nabla c \otimes \nabla c)]\cdot \mathbf{u} \bigg\} \delta V + \int_\Gamma \kappa \nabla c\cdot \mathbf n \frac{\partial c}{\partial t} \delta \Gamma. \label{eq:40-7} \end{align}
  The first volume term, together with Eq.~\eqref{eq:20-2}, yields the Cahn--Hilliard equation. The second volume term generates the Korteweg stress associated with the composition gradient, thereby establishing a connection between the Lagrangian kinetic Gibbs--Duhem formulation and the conventional Eulerian description of diffuse-interface mechanics. The resulting form of the Korteweg stress tensor, proportional to $\nabla c\otimes\nabla c$, is consistent in part with the formulation of Ref.~\cite{anderson1998diffuse}. Unlike the classical Korteweg stress tensor~\cite{korteweg1901forme}, the pressure contribution is explicitly determined in the present formulation through Eq.~\eqref{eq:37-8}. Finally, the boundary term in Eq.~\eqref{eq:40-7} provides a thermodynamic contribution associated with chemical wetting effects~\cite{wang2023thermodynamically}, thereby establishing a direct connection between the gradient-energy formulation and the thermodynamics of wetting.

\section{Electromagnetic field and its coupling with mass field}

\subsection{Free energy functional and the Gibbs-Duhem relation}
The free-energy functional for the coupled electromagnetic and material system is written as
\begin{align}
\notag
\mathcal E
=&\int_V
\left(
\rho_e\psi
+\mathbf D\cdot\mathbf E
+\rho_b\phi
+\mathbf B\cdot\mathbf H
+\rho_e\mathbf u^2
+\rho_p\mathbf c^2
+p
\right)\delta V,\\
=&\int_V
\left(
\rho_e\psi
+\varepsilon\mathbf E^2
+\rho_b\phi
+\chi\mathbf H^2
+(\mathbf B\times\mathbf D)\cdot\mathbf u
+\rho_e\mathbf u^2
+\rho_p\mathbf c^2
+p_m+p_c
\right)\delta V.
\end{align}
Here, $\mathbf E$ and $\mathbf H$ denote the electric and magnetic field variables, respectively, while $\mathbf D$ and $\mathbf B$ are their corresponding constitutive fields~\cite{ogden1997nonlinear},
\begin{equation}
\mathbf D=\varepsilon\mathbf E,
\qquad
\mathbf B=\chi\mathbf H,
\end{equation}
where $\varepsilon$ and $\chi$ are the electric permittivity and magnetic permeability, respectively. Within the present formulation, the electromagnetic field variables are associated with the local transport of electromagnetic energy, while $\mathbf D$ and $\mathbf B$ provide the corresponding constitutive responses. The quantities $\rho_e$ and $\rho_b$ denote the electric and magnetic source densities, respectively, and $\rho_p$ represents the effective mass density associated with electromagnetic-field transport. The scalar fields $\psi$ and $\phi$ are the corresponding potentials of the static electromagnetic fields.

The material transport contribution $\rho_e\mathbf u^2$ can alternatively be expressed through the electromagnetic momentum associated with the Poynting-vector representation~\cite{mcdonald2026poynting,tobar2022poynting},
\begin{equation}
\rho_e\mathbf u\cdot\mathbf u
=(\mathbf B\times\mathbf D)\cdot\mathbf u.
\end{equation}

As discussed in Section IIIB, the kinetic energies of both the photon and the massive particles are not Galilean/Lorentz invariant. As pointed out by Kovetz~\cite{kovetz2000electromagnetic}, the electric field $\mathbf{E}$ composed of photon particles is likewise not Galilean/Lorentz invariant. When we write $\varepsilon_0\mathbf{E}^2=\rho_p\mathbf{c}^2$, 
these two statements are mutually consistent.
To overcome this problem, we introduce the pressure $p$ acting as a Lagrange multiplier enforcing the constraint associated with the constant volume element $\delta V$ containing the electromagnetic and material fields. We separate this multiplier into material and electromagnetic contributions,
\begin{equation}
p=p_m+p_c,
\end{equation}
where $p_m$ is associated with the charge-carrying material and $p_c$ with electromagnetic-field transport.

Writing the free-energy functional in this form makes it possible to separate its differential into dissipative and nondissipative contributions. Its total differential is
\begin{align}
\notag
\mathrm d\mathcal E
=\int_V&\Big\{
\left[
\rho_e\mathrm d\psi
+\varepsilon\mathbf E\cdot\mathrm d\mathbf E
+\rho_b\mathrm d\phi
+\chi\mathbf H\cdot\mathrm d\mathbf H
+\rho_e\mathbf u\cdot\mathrm d\mathbf u
+\rho_p\mathbf c\cdot\mathrm d\mathbf c
\right]
\\
+
&\left[
\psi\,\mathrm d\rho_e
+\mathbf E\cdot\mathrm d(\varepsilon\mathbf E)
+\phi\,\mathrm d\rho_b
+\mathbf H\cdot\mathrm d(\chi\mathbf H)
+\mathbf u\cdot\mathrm d(\rho_e\mathbf u)
+\mathbf c\cdot\mathrm d(\rho_p\mathbf c)
+\mathrm dp
\right]
\Big\}\delta V.
\label{eq:15-5}
\end{align}
The first bracket contains variations of the field and transport variables and represents the dissipative contribution associated with entropy production. In particular, the corresponding variational conditions lead to the Gauss laws for the electromagnetic fields, which are derived in the following section. The second bracket contains the remaining nondissipative contribution. 
Following the construction introduced for material transport, this contribution defines the kinetic Gibbs--Duhem relation for the coupled electromagnetic system.

Because the electromagnetic field has both a propagating component and a material-coupled component, the Gibbs--Duhem contribution separates naturally into two relations. The first describes the electromagnetic-field transport,
\begin{equation}
\label{eq:41-0}
\boxed{
\mathbf c\cdot\mathrm d(\rho_p\mathbf c)
+\mathrm dp_c=0,
}
\end{equation}
which is the electromagnetic analogue of the kinetic Gibbs--Duhem relation for a transported particle.
The second incorporates the coupling among charge, electric and magnetic fields, and material transport,
\begin{equation}
\label{eq:41}
\boxed{
\psi\,\mathrm d\rho_e
+\mathbf E\cdot\mathrm d(\varepsilon\mathbf E)
+\phi\,\mathrm d\rho_b
+\mathbf H\cdot\mathrm d(\chi\mathbf H)
+\mathbf u\cdot\mathrm d(\rho_e\mathbf u)
+\mathrm dp_m=0.
}
\end{equation}
Using the relation between material transport momentum and the electromagnetic momentum representation, this equation can equivalently be written as
\begin{equation}
\boxed{
\psi\,\mathrm d\rho_e
+\mathbf E\cdot\mathrm d(\varepsilon\mathbf E)
+\phi\,\mathrm d\rho_b
+\mathbf H\cdot\mathrm d(\chi\mathbf H)
+\mathbf u\cdot\mathrm d(\mathbf B\times\mathbf D)
+\mathrm dp_m=0.
}
\end{equation}
Thus, Eqs.~\eqref{eq:41-0} and \eqref{eq:41} provide the thermodynamic constraints associated with electromagnetic-field propagation and electromagnetic--mechanical coupling, respectively.

Integrating these Gibbs--Duhem relations gives the corresponding pressure contributions,
\begin{equation}
-p_c=\rho_p\mathbf c^2,
\ \text{or}\
E=\rho_p\delta V\,\mathbf c^2=m\mathbf c^2,
\end{equation}
for the electromagnetic-field transport, and
\begin{equation}
-p_m
=\rho_e\mathbf u^2
+\varepsilon\mathbf E^2
+\chi\mathbf H^2
+\rho_e\psi
+\rho_b\phi.
\label{eq:50-89}
\end{equation}
These relations show that the Lagrange multipliers $p_c$ and $p_m$ absorb the frame-dependent kinetic and field contributions into the thermodynamic constraint. Substitution into the free-energy functional therefore yields an energy expression consistent with the Galilean and Lorentz transformation properties considered in the present formulation.

The kinetic Gibbs--Duhem relation also gives the corresponding momentum evolution equation. Taking its spatial differential yields
\begin{equation}
\frac{\mathrm d(\rho_e\mathbf u)}{\mathrm dt}
=
-\nabla p_m
-\mathbf E\cdot\nabla(\varepsilon\mathbf E)
-\mathbf H\cdot\nabla(\chi\mathbf H)
-\psi\nabla\rho_e
-\phi\nabla\rho_b.
\label{eq:50-222}
\end{equation}
Using the electromagnetic representation of the transport momentum, this equation can equivalently be expressed as
\begin{equation}
\frac{\mathrm d(\mathbf B\times\mathbf D)}{\mathrm dt}
=
-\nabla p_m
-\mathbf E\cdot\nabla(\varepsilon\mathbf E)
-\mathbf H\cdot\nabla(\chi\mathbf H)
-\psi\nabla\rho_e
-\phi\nabla\rho_b.
\end{equation}
The resulting equation provides the electromagnetic--mechanical momentum balance from the same Gibbs--Duhem structure that governs chemical and material transport. In this sense, the electromagnetic field equations and the associated momentum balance are not introduced as independent principles, but arise from the separation of dissipative and nondissipative contributions in the free-energy differential.

\subsection{Maxwell equations in the Eulerian framework}

The classical Maxwell equations are formulated in the Eulerian framework, in which the time derivative does not explicitly include the convective contribution associated with the material velocity~\cite{kovetz2000electromagnetic}. Consequently, the relativistic transformation properties associated with the propagation of light are not explicitly represented in the conventional Maxwell formulation.
The dissipation law in the Eulerian framework without convection velocity  reads
\begin{equation}
 \frac{\partial \mathcal E}{\partial t}=\int_V \bigg(\rho_e\frac{\partial \psi}{\partial t}+ \varepsilon \mathbf{E}\cdot \frac{\partial \mathbf{E}}{\partial t}+ \rho_b  \frac{\partial \phi}{\partial t} +\chi \mathbf{H}\cdot \frac{\partial \mathbf{H}}{\partial t}+ \rho_e \mathbf{u}\cdot \frac{\partial \mathbf{u}}{\partial t}\bigg) \delta V.
 \label{eq:45-77}
\end{equation}
In the case of statics for  electromagnetic field, the electric-magnetic field can be expressed as   $\mathbf{E}=\nabla \psi$ and $\mathbf H=\nabla \phi$.
Applying integration by parts to the second and fourth terms in Eq.~\eqref{eq:45-77}, we obtain the following dissipation laws 
\begin{equation}
 \frac{\partial \mathcal E}{\partial t}=\int_V [
 (\rho_e-\nabla\cdot \varepsilon \nabla \psi) \frac{\partial \psi}{\partial t} + (\rho_b- \nabla \cdot \chi \nabla \phi)\frac{\partial \phi}{\partial t}] \delta V 
 +\int_\Gamma  (\varepsilon \mathbf{E}\cdot \mathbf{n} \frac{\partial \psi}{\partial  t} + \chi\mathbf{H}\cdot \mathbf{n} \frac{\partial \phi}{\partial  t})  \delta \Gamma.
\end{equation}
The first and second terms lead to Gauss's law
\begin{align}
 \rho_e=\nabla\cdot \varepsilon \nabla \psi,\\
 \rho_b= \nabla \cdot \chi \nabla \phi,
\end{align}
for electrical field and magnetic field, respectively.
The boundary term describes the refraction and reflection of electromagnetic field at the interface, which is out of the scope of the current work.

The momentum balance equation in the Eulerian framework
 reads
\begin{equation}
 \frac{ \partial (\mathbf B \times \mathbf{D})}{\partial t}=-\nabla p_m-\mathbf{E} \cdot \nabla (\varepsilon \mathbf{E})-  \mathbf{H} \cdot \nabla (\chi \mathbf{H})-\psi\nabla \rho_e-\phi\nabla \rho_b.
 \label{eq:35-9}
\end{equation}

The left hand side of Eq.~\eqref{eq:35-9}  can be expanded as 
\begin{equation}
 \frac{\partial(\bold B\times\bold D)}{\partial t}=\bold B\times \frac{\partial \bold D}{\partial t}
 -\bold D\times \frac{\partial \bold B}{\partial t}.
 \label{eq:83-9}
\end{equation}

The second and third terms on the right hand side of Eq.~\eqref{eq:35-9} 
can be expanded as 
\begin{align}
\notag
-\mathbf E\cdot\nabla \mathbf D
&=-\nabla(\mathbf E\cdot\mathbf D)
+
(\mathbf D\cdot\nabla)\mathbf E
+
\mathbf E\times(\nabla\times\mathbf D)
+
\mathbf D\times(\nabla\times\mathbf E)\\
 \label{eq:83-10}
&=-\nabla(\mathbf E\cdot\mathbf D)
+
(\mathbf D\cdot\nabla)\mathbf E
+
\mathbf D\times(\nabla\times\mathbf E),\ \text{and} \\
\notag
-\mathbf H\cdot\nabla \mathbf B
&=-\nabla(\mathbf H\cdot\mathbf B)
+
(\mathbf B\cdot\nabla)\mathbf H
+
\mathbf H\times(\nabla\times\mathbf B)
+
\mathbf B\times(\nabla\times\mathbf H)\\
&=-\nabla(\mathbf H\cdot\mathbf B)+(\mathbf B\cdot\nabla)\mathbf H+\mathbf B\times(\nabla\times\mathbf H)+\mathbf J\times \mathbf B+\mathbf B\times \mathbf J.
 \label{eq:83-11}
\end{align}
In the above derivation, we have added a zero term, $\mathbf J\times \mathbf B+\mathbf B\times \mathbf J=\mathbf{ 0}$
and used the conditions: $\nabla \times \mathbf{D}=\mathbf{ 0}$ and $\nabla \times \mathbf{B}=\mathbf{ 0}$.
The derivation of the last two conditions is presented in Section VD.

Collecting the $\mathbf{D}\times$ terms in Eq.~\eqref{eq:83-9} and Eq.~\eqref{eq:83-10},
 we obtain the Maxwell-Faraday equation
 \begin{equation}
  \frac{\partial \mathbf B}{\partial  t}=-\nabla \times \mathbf{E}.
 \end{equation}

 Collecting the $\mathbf{B}\times$ terms Eq.~\eqref{eq:83-9} and Eq.~\eqref{eq:83-11},
 we obtain the Maxwell-Ampere equation
 \begin{equation}
  \frac{\partial \mathbf D}{\partial  t}=\mathbf{J}+ \nabla\times \mathbf{H}.
 \end{equation}
 
\subsection{Generalization of the Maxwell equations in the Lagrangian framework and superconductivity} 
 
As mentioned, in the classic  Maxwell equation, the convective part has not been considered in the time derivative.
To account for the convective effect, we write the dissipation law for the electrical field as 
\begin{equation}
 \int_V [\rho_e\frac{\mathrm d \psi }{\mathrm d t}+\varepsilon \mathbf{E} \cdot \frac{\mathrm d \mathbf{E} }{\mathrm  d t}] \delta V
 = \int_V [\rho_e\frac{\mathrm d \psi }{\mathrm d t}+\varepsilon \nabla \psi \cdot \frac{\mathrm d \nabla \psi }{\mathrm d t} ] \delta V
=\int_V[\rho_e\frac{\mathrm d \psi }{\mathrm d t}+ \varepsilon \nabla \psi \cdot (\frac{\partial  \nabla \psi }{\partial t}+ \mathbf{u}\cdot \nabla \nabla \psi) ] \delta V.
 \end{equation}
Applying integration by parts to the first term and combining like terms,
we obtain the Maxwell stress tensor and the Gauss's law
  \begin{equation}
 \int_V \bigg[ (\rho_e -\nabla\cdot \varepsilon\nabla \psi) \frac{\mathrm d \psi }{\mathrm d t}+ \nabla \cdot (\varepsilon \nabla \psi \otimes \nabla \psi )\cdot \mathbf{u} \bigg] \delta V.
 \label{eq:64-8}
 \end{equation}
 Repeating the same calculation, we obtain the energy dissipation
 for the magnetic field as 
  \begin{equation}
 \int_V \bigg[(\rho_b -\nabla\cdot \chi\nabla \phi) \frac{\mathrm d \phi }{\mathrm d t}+ \nabla \cdot (\chi \nabla \phi \otimes \nabla \phi )\cdot \mathbf{u}\bigg] \delta V.
 \label{eq:64-9}
 \end{equation}
 From the above two equations,
 we obtain Gauss's law 
 as well as  the  Maxwell stress tensor for the  static electromagnetic field as
 \begin{equation}
  \mathcal M=-\varepsilon \nabla \psi \otimes \nabla \psi-\chi \nabla \phi \otimes \nabla \phi.
 \end{equation}

Putting the Maxwell stress tensor,
$-\nabla\cdot \mathcal M$
into the momentum evolution equation
and using the expression for the pressure [Eq.~\eqref{eq:50-89}],
we obtain the Maxwell equations in the  Lagrangian framework 
 \begin{align}
  \frac{\mathrm d \mathbf B}{\mathrm d  t}&=-\nabla \times \mathbf{E},\\
  \frac{\mathrm d\mathbf D}{\mathrm d  t}&=\mathbf{J}+ \nabla\times \mathbf{H}.
 \end{align}
 The remaining terms in the momentum evolution equation are 
 \begin{align}
 \boxed{
 \nabla (\rho_e \mathbf{ u}^2)+ \mathbf J\times \mathbf B =\mathbf{0},
 }
 \label{eq:69}
 \end{align}
which can be used to interpret the superconductivity of both type-I and type-II superconductors.

For a type-I superconductor, the magnetic induction vanishes inside the conductor, $\mathbf B=\mathbf 0$, corresponding to the Meissner effect. Within the present thermodynamic framework in the Lagrangian description, the electron transport is governed by the dissipation law
\begin{equation}
\int_V \rho_e\mathbf{u}\cdot\frac{\mathrm d\mathbf{u}}{\mathrm dt} \delta V\leq 0.
\end{equation}
The corresponding evolution equation is
\begin{equation}
\frac{\mathrm d\mathbf{u}}{\mathrm dt}
=\nabla\cdot D_u\nabla(\rho_e\mathbf{u}).
\label{eq:70-999}
\end{equation}
When the electron transport momentum $\rho_e\mathbf u$ is spatially uniform, the right-hand side vanishes, and the transport velocity evolves without dissipation. Consequently, the electron kinetic energy remains constant during transport, providing the thermodynamic condition for nondissipative transport associated with superconductivity.

For a type-II superconductor, superconducting and normal regions coexist. In the normal regions, the electric current vanishes, and hence $\mathbf{J}\times\mathbf{B}=\mathbf{0}$, regardless of whether $\mathbf{B}$ is zero or finite. Under this condition, there is no dissipation of the electrons' kinetic energy during transport. By contrast, when $\mathbf{J}\times\mathbf{B}\neq\mathbf{0}$, a dissipative contribution arises, corresponding to a finite resistance in the normal-conducting state.

A microscopic interpretation of these conditions requires expressing the electron kinetic energy, $\rho_e\mathbf u^2$, in terms of the quantum-mechanical wave function. Such an extension may clarify the microscopic origin of the nondissipative conditions, including $\mathbf J \times\mathbf B=\mathbf 0$. Developing this connection between the present thermodynamic framework and quantum mechanics will be considered in future work.

\subsection{Parity asymmetry in the Lagrangian framework}

To account for the convective effect associated with electron transport, the electric field is represented in the Lagrangian framework as
\begin{equation}
\mathbf{E}=\nabla \psi+\nabla\times\boldsymbol{\varphi}.
\label{eq:60-88}
\end{equation}
The corresponding dissipation law becomes
\begin{align} 
&\int_V \bigg[\rho_e \frac{\mathrm d \psi}{\mathrm d t}+\varepsilon (\nabla \psi+ \nabla \times \boldsymbol\varphi) \cdot \frac{\mathrm d (\nabla \psi+\nabla \times \boldsymbol\varphi) }{\mathrm d t} \bigg] \delta V\\ &=\int_V \bigg\{[\rho_e-\nabla \cdot \mathbf D ]\frac{\mathrm{d} \psi}{\mathrm{d}t} + (\nabla\times \bold D)\cdot \frac{\partial \boldsymbol \psi_1}{\partial t } + [ \nabla \cdot (\mathbf{D}\otimes \mathbf E)+ \rho_e \nabla \times \boldsymbol\varphi]\cdot \mathbf{u} + \varepsilon^\prime \bold E^2\frac{\mathrm{d} c}{\mathrm{d}t } \bigg\} \delta V.
\end{align}
The first term yields the generalized Gauss law. The second term gives the additional equilibrium condition
\begin{equation}
\nabla\times\mathbf D=\mathbf 0.
\end{equation}
The third term contains the Maxwell stress contribution for a nonstatic electromagnetic field. The last term represents the coupling between the electric field and the mass field and consequently contributes to the chemical potential. In addition, the Lagrangian formulation introduces the force term
$\rho_e\nabla\times\boldsymbol{\varphi}$,
which has no counterpart in the conventional Eulerian formulation. The corresponding momentum equation can be written as
\begin{equation}
\frac{\mathrm d(\rho_e\mathbf u)}{\mathrm dt}
=-\rho_e\nabla\times\boldsymbol{\varphi}
+\text{other terms}.
\end{equation}
The appearance of this curl contribution distinguishes the transformation properties of the Lagrangian formulation and leads to a parity-asymmetric contribution to the momentum balance.

An analogous calculation can be performed for the magnetic field. We write
\begin{equation}
\mathbf H=\nabla\phi+\nabla\times\mathbf A,
\end{equation}
which gives the dissipation law
\begin{align}
&\int_V \bigg[\rho_b\frac{\mathrm d \phi }{\mathrm d t}+\varepsilon (\nabla \phi+ \nabla \times \bold A)\cdot \frac{\mathrm d (\nabla \phi+\nabla \times \bold A) }{\mathrm d t} \bigg] \delta V\\
&=\int_V \bigg\{(\rho_b-\nabla \cdot \mathbf B )\frac{\mathrm{d} \phi}{\mathrm{d}t} + (\nabla\times \bold B)\cdot \frac{\partial \bold A}{\partial t } + [ \nabla \cdot (\mathbf{B}\otimes \mathbf H)+ \rho_b \nabla \times \bold A]\cdot \mathbf{u} + \chi^\prime \bold H^2\frac{\mathrm{d} c}{\mathrm{d}t } \bigg\} \delta V.
\end{align}
The first term gives Gauss's law for the magnetic field. The second term yields the additional equilibrium condition
\begin{equation}
\nabla\times\mathbf B=\mathbf 0.
\end{equation}
The third term contains the Maxwell stress contribution associated with the magnetic field, while the last term represents the coupling between the magnetic field and the mass field through the chemical potential. Because the magnetic charge density vanishes, $\rho_b=0$, the corresponding curl force term $\rho_b\nabla\times\mathbf A$ is absent.

The physical meaning of the curl contribution in Eq.~\eqref{eq:60-88} can be further clarified by substituting Eq.~\eqref{eq:60-88} into the Maxwell-Faraday equation. In vacuum, with
\begin{equation}
\mathbf H=\nabla\times\mathbf A,
\end{equation}
the Maxwell-Faraday equation gives
\begin{equation}
\chi_0\frac{\partial\mathbf A}{\partial t}
=-\nabla\times\boldsymbol{\varphi}.
\end{equation}
Thus, the curl component of the electric field is directly related to the time evolution of the magnetic vector potential. This relation provides a Lagrangian interpretation of the curl contribution in Eq.~\eqref{eq:60-88} and connects the additional force term in the momentum balance with electromagnetic-field transport.

\subsection{Entropy origin for the velocity evolution and a common thermodynamic structure for the speed of sound and speed of light}

As demonstrated above, both mass and electromagnetic fields admit a common thermodynamic structure for momentum evolution. For the mass field, this structure leads to Newton's law, whereas for the electromagnetic field it leads to Maxwell's equations. The physical significance of this common thermodynamic structure is that it ensures the Galilean invariance of the system free energy.

Conventionally, in the Eulerian description with convection, the momentum balance is written as
\begin{equation}
\frac{\partial (\rho \mathbf u)}{\partial t}
+\nabla \cdot \mathbf{J}_u=\mathbf{0},
\label{eq:87-7}
\end{equation}
where the momentum-flux tensor is assumed to be
\begin{equation}
\mathbf{J}_u=\rho \mathbf u\otimes\mathbf u+p\mathbf I.
\end{equation}
Eq.~\eqref{eq:87-7} corresponds to the Boltzmann transport equation for the first moment, $\rho\mathbf u^1=\rho\mathbf u$. Expanding the divergence of the momentum-flux tensor gives
\begin{equation}
\rho\left(
\frac{\partial\mathbf u}{\partial t}
+\mathbf u\cdot\nabla\mathbf u
\right)
+\left[
\frac{\partial\rho}{\partial t}
+\nabla\cdot(\rho\mathbf u)
\right]\mathbf u
=-\nabla p.
\label{eq:87-8}
\end{equation}

For the zeroth moment, $\rho\mathbf u^0=\rho$, the Boltzmann transport equation reduces to the continuity equation,
\begin{equation}
\frac{\partial\rho}{\partial t}
+\nabla\cdot(\rho\mathbf u)=0.
\label{eq:88-7}
\end{equation}
The same Eulerian framework can be applied to electron transport, with the mass density $\rho$ replaced by the electron density $\rho_e$.

There are two limitations to this conventional formulation. First, in the presence of diffusion, as in electron transport, the continuity equation acquires an additional diffusive contribution, which may be written, for a constant diffusion coefficient, as $D_\rho\nabla^2\rho$. In the present formulation, this diffusive contribution is associated with electrical resistance. Substitution of the resulting density evolution into Eq.~\eqref{eq:87-8} generates the term
\begin{equation}
\left[
\frac{\partial\rho}{\partial t}
+\nabla\cdot(\rho\mathbf u)
\right]\mathbf u,
\end{equation}
which is not Galilean invariant because $\mathbf u$ changes under a Galilean transformation.

We avoid this difficulty by assigning the acceleration an entropic origin. The essential point is to formulate the momentum evolution first in the Lagrangian description, rather than deriving it from an assumed Eulerian momentum flux through Gauss's theorem. The corresponding Lagrangian form is
\begin{equation}
\frac{\partial(\rho\mathbf u)}{\partial t}
+\mathbf u\cdot\nabla(\rho\mathbf u),
\end{equation}
in place of the Eulerian flux form
\begin{equation}
\nabla\cdot(\rho\mathbf u\otimes\mathbf u).
\end{equation}
The velocity is then governed by an independent evolution equation, Eq.~\eqref{eq:70-999}. The density evolution is obtained by combining the momentum evolution equation [Eq.~\eqref{eq:50-222}] with the velocity-dissipation equation [Eq.~\eqref{eq:70-999}]. In this formulation, the resulting equations preserve Galilean invariance.

It is important to note that density is not, in general, a conserved quantity. Processes such as evaporation and condensation at room temperature, as well as Bose--Einstein condensation at low temperatures, can lead to changes in the density following the Lagrangian trajectory.
In contrast, in the absence of particle creation or annihilation, the total number of molecules is the conserved quantity. Consequently, variations in density may arise from changes in the spatial distribution, volume, or phase of the molecular system, even when the total number of molecules remains constant.

The second limitation concerns the conventional treatment of viscosity. In the Eulerian momentum balance [Eq.~\eqref{eq:87-7}], the viscous stress tensor is introduced phenomenologically as an additional mechanical force. In electron transport, the analogous dissipative contribution is associated with electrical resistance. In the present formulation, both viscous dissipation and electrical resistance are instead attributed to an entropic contribution, with the mechanical and entropic forces constrained by the Gibbs--Duhem relation. This structure is incorporated through the entropy evolution equation for the velocity, Eq.~\eqref{eq:70-999}. The resulting velocity dynamics provides a common framework for describing both superconducting behavior and finite electrical resistance.

Importantly, assigning an entropic origin to the velocity evolution does not contradict Newtonian mechanics. In the constant-density limit, Newton's law is recovered from the kinetic Gibbs--Duhem relation. Thus, Newtonian dynamics emerges as a limiting case of the more general thermodynamic formulation.

The density evolution can  be obtained by combining the velocity-dissipation equation [Eq.~\eqref{eq:70-999}] with the momentum evolution equation [Eq.~\eqref{eq:50-222}]. Under isentropic conditions, the momentum balance for both the mass and electromagnetic fields takes the common form
\begin{equation}
\mathbf{u}\frac{\mathrm d\rho}{\mathrm dt}=-\nabla p.
\end{equation}
This relation provides a unified definition of the characteristic propagation velocity. Introducing the thermodynamic pressure~\footnote{
As a Lagrange multiplier, $p$ may appear with either a `+'' or `-'' sign, depending on the choice of the normal-vector orientation. The two signs correspond to outward- and inward-pointing normal vectors, respectively.}
\begin{equation}
\mathcal P=-p,
\end{equation}
we obtain
\begin{equation}
u^2=\frac{\mathrm d\mathcal P}{\mathrm d\rho}.
\end{equation}
For the mass field, this relation identifies $u$ with the speed of sound, whereas for the electromagnetic field it identifies the corresponding characteristic velocity with the speed of light.

\section{Conclusion}

We have developed a unified thermodynamic framework for chemical, mechanical, and electromagnetic transport based on a Lagrangian formulation of free-energy dissipation. The central result is an extension of the Gibbs--Duhem relation to include the kinetic free energy associated with material and electromagnetic transport. This kinetic Gibbs--Duhem relation provides a common thermodynamic basis for chemical diffusion, momentum transport, electromagnetic-field dynamics, and their coupling.

For chemical systems, the formulation yields coupled diffusion and momentum transport. The kinetic Gibbs--Duhem relation,
\begin{equation}
\tau\mathrm d\mu_\tau+c\mathrm d\mu+\mathrm dp
+\mathbf u\cdot\mathrm d(\rho\mathbf u)=0,
\end{equation}
gives the momentum balance
\begin{equation}
\frac{\mathrm d(\rho\mathbf u)}{\mathrm dt}
=-\nabla p-c\nabla\mu-\tau\nabla\mu_\tau.
\end{equation}
For nonuniform systems, the gradient contribution leads to
\begin{equation}
\frac{\mathrm d(\rho\mathbf u)}{\mathrm dt}
=-\nabla p-\nabla c\cdot\nabla(\kappa\nabla c),
\end{equation}
establishing a direct connection between the Lagrangian thermodynamic formulation, Cahn--Hilliard dynamics, and the Korteweg stress associated with composition gradients.

For electromagnetic systems, the corresponding Gibbs--Duhem relations are
\begin{equation}
\mathbf c\cdot\mathrm d(\rho_p\mathbf c)+\mathrm dp_c=0,
\end{equation}
and
\begin{equation}
\psi\mathrm d\rho_e
+\mathbf E\cdot\mathrm d(\varepsilon\mathbf E)
+\phi\mathrm d\rho_b
+\mathbf H\cdot\mathrm d(\chi\mathbf H)
+\mathbf u\cdot\mathrm d(\rho_e\mathbf u)
+\mathrm dp_m=0.
\end{equation}
These relations provide a thermodynamic basis for the associated mass, energy, momentum, and Maxwell equations. The same framework also describes nondissipative superconducting transport. For type-I superconductors, the Meissner condition $\mathbf B=\mathbf0$ together with uniform transport momentum yields nondissipative transport, whereas for type-II superconductors,
$\mathbf J\times\mathbf B=\mathbf0$
identifies the nondissipative vortex-transport condition.

The Lagrangian formulation further retains coupling terms associated with curl components of the electromagnetic fields. In particular,
\begin{equation}
\mathbf E=\nabla\psi+\nabla\times\boldsymbol\varphi
\end{equation}
introduces the additional force $\rho_e\nabla\times\boldsymbol\varphi$ into the momentum balance. This term distinguish the Lagrangian formulation from the conventional Eulerian description and provide additional contributions associated with convective electromagnetic transport.

Finally, the formulation reveals a common thermodynamic structure for mechanical and electromagnetic propagation. Under isentropic conditions, the momentum balance gives
\begin{equation}
u^2=\frac{\mathrm d\mathcal P}{\mathrm d\rho},
\end{equation}
where the constitutive relation determines the corresponding propagation velocity. Thus, acoustic and electromagnetic propagation can be described within a common thermodynamic framework.

Overall, the kinetic Gibbs--Duhem relation provides a unified description of chemical diffusion, mechanical transport, electromagnetic dynamics, and superconducting transport. By treating dissipative and nondissipative contributions within a common Lagrangian free-energy formulation, the framework provides a systematic basis for thermodynamic modeling of coupled chemical--electromagnetic systems.

An extension of the present framework to systems undergoing large deformation provides a natural direction for future work. When spatial variations of the transport velocity become significant, the velocity gradient $\nabla\mathbf u$ must be retained explicitly in the thermodynamic description. In the electromagnetic sector, the corresponding spatial variation of the electric field, $\nabla\mathbf E$, should likewise be considered, particularly when electromagnetic fields are coupled to strongly deforming or spatially heterogeneous media. The extension can be formulated systematically by decomposing the energy density into dissipative and nondissipative contributions and treating the associated gradients within the same Lagrangian variational framework. In this formulation, $\nabla\mathbf u$ and $\nabla\mathbf E$ would generate additional constitutive and momentum-coupling terms while preserving the distinction between evolution equations arising from dissipation and constraints arising from the nondissipative part. Such an extension could provide a unified thermodynamic description of coupled electromagnetic transport and finite-deformation mechanics and is left as an outlook for future investigation.

\acknowledgments

I am grateful to Prof. Dr. B. Nestler, L. Bi, and R. Yin for valuable discussions. I also thank Prof. Howard Stone (Princeton University) for helpful discussions.

\end{document}